\documentclass[epj]{svjour}

\usepackage{graphicx}
\usepackage{fancyhdr}
\usepackage{amsmath,amssymb,bbm,mathrsfs}

\newlength{\textlength}
\newlength{\overlinelength}
\newcommand{\ovl}[2][.55]{\settowidth{\textlength}{$#2$}
  \setlength{\overlinelength}{0.1pt}
  \addtolength{\overlinelength}{0.75\textlength}
  \makebox[\textlength][s]{$#2$} \hspace{-.55\textlength}
  \hspace{-\overlinelength}\hspace{#1\overlinelength}
  \overline{\makebox[\overlinelength][s]{\vphantom{$#2$}}}
  \hspace{-#1\overlinelength}\hspace{.55\textlength}}

\DeclareMathOperator{\diag}{diag}
\DeclareMathOperator{\im}{Im}

\newcommand{\abs}[1]{\left| #1\right|} 
\newcommand{\bm}{\overline{\mu}}       
\newcommand{\rt}{\widetilde{\rho}}     
\newcommand{\rb}{\overline{\rho}}      
\newcommand{\tm}{\widetilde{\mu}}      
\newcommand{\tM}{\widetilde{M}}        
\newcommand{\VEVsmall}[1]{\langle #1\rangle}       

\def\jphg#1#2#3{J.~Phys.~{\bf G {#1}} ({#2}) #3}

\def\plb#1#2#3{Phys.~Lett.~{\bf B {#1}} ({#2}) #3}

\def\prd#1#2#3{Phys.~Rev.~{\bf D {#1}} ({#2}) #3}

\def\mytitle{My title} 
\def\myauthors{My name}  
\def\mytype{My type of session}
\def\mysession{My session}

\def\mytitle{Neutrino Masses and Mixings from Quark Mass
  Hierarchies}
\def\myauthors{S\"oren Wiesenfeldt}
\def\mytype{Contributed Talk}    
\def\mysession{Flavor Physics}

\begin{document}

\title{Neutrino Masses and Mixings from Quark Mass Hierarchies}

\author{S\"oren Wiesenfeldt\thanks{\emph{Present address:} Institut
    f\"ur Theoretische Teilchenphysik, Universit\"at Karlsruhe, 76128
    Karlsruhe.}
}%

\institute{Department of Physics, University of Illinois at
  Urbana-Champaign, 1110 West Green Street, Urbana, IL 61801, USA 
}%

\date{}

\abstract{%
  In SO(10) models with a vectorial fourth generation of down quarks and
  leptons, the structure of the neutrino Majorana and Dirac mass
  matrices generically coincide with those of up- and down-quarks,
  respectively.  Then the small neutrino mass hierarchy follows from the
  mismatch of the up and down quark mass hierarchies and we can
  accommodate naturally a normal hierarchy.  The effective CP violating
  phases in the quark sector, neutrino oscillations and leptogenesis are
  unrelated.
}

\maketitle

\section{Introduction}\label{intro}

The symmetries and the particle content of the standard model (SM) point
towards grand unified theories (GUTs) as an underlying theory.  In
particular, SO(10) is a very attractive candidate: All quarks and
leptons of one generation are unified in a single multiplet,
$\mathsf{16} = \left(Q,u^c,d^c,L,\nu^c,e^c\right)$, where $\nu^c$ can be
identified with the right-handed neutrino; it is anomaly free; and it
contains a gauged $\text{U(1)}_{B-L}$ subgroup.  Thus, SO(10) predicts
massive neutrinos and it is striking that the $B-L$ breaking scale,
$M_{B-L}$, is close to the scale, where the SM gauge couplings converge,
$M_\text{GUT}$.

The masses and mixings of neutrinos are very different from those of
quarks.  While the quark masses are strongly hierarchical and their
mixing is small, the neutrino mixing is close to tribimaximal and their
masses can be degenerate.  Even if they are hierarchical, the mass ratio
of the heavier neutrino masses is constrained as $m_2/m_3 > \sqrt{\Delta
  m_\text{sol}^2/\Delta m_\text{atm}^2} \simeq 0.2$.

In GUT models, fermion masses arise from a few Yukawa couplings, which
implies relations among the fermion mass matrices.  In most models, the
Dirac mass matrices of up quarks and neutrinos, $m^u$ and $m^\nu$, as
well as those of down quarks and charged leptons, $m^d$ and $m^e$, are
strongly related,
\begin{align}
  \label{eq:mass-rel}
  m^u \sim m^\nu \; , \qquad m^d \sim m^e \;.
\end{align}
Whether or not the Majorana mass matrix of the right-handed neutrino,
$m^N$, is related to the other matrices, depends on the specific
breaking mechanism.

The pattern (\ref{eq:mass-rel}), however, is no longer valid if we have
additional heavy matter.  Consider a model with three spinorial and one
vectorial matter fields, $\mathsf{16}_{1,2,3}$ and $\mathsf{10}_M$
\cite{4gen}.  The presence of $\mathsf{10}_M$ adds a fourth generation
of down quarks and leptons.  Then $m^u$ is a $3\times 3$ matrix, whereas
$m^d$, $m^e$ and $m^\nu$ are $4\times 4$,
\begin{align}
  m^u & \sim
  \begin{pmatrix}
    \mathsf{M}_{\alpha\beta}^{(3\times 3)} \vphantom{\dfrac{M}{M}}
  \end{pmatrix}
  , & m^{d,e,\nu} & \sim \left(
    \begin{array}{c|c}
      \mathsf{M}_{\alpha\beta}^{(3\times 3)} \vphantom{\dfrac{M}{M}}
      & \mathsf{M}_{\alpha4} \cr 
      \hline
      \mathsf{M}_{4\alpha} & \mathsf{M}_{44}
    \end{array}
  \right) .
\end{align}
The fourth generation acquires GUT-scale masses when the GUT symmetry is
broken.  If $\mathsf{M}_{4\alpha}$ or $\mathsf{M}_{\alpha4}$ are
comparable to $\mathsf{M}_{44}$, the mixing among left-handed (LH) or
right-handed (RH) states is large and the effective $3\times 3$ mass
matrix is much different from $m^u \sim
\mathsf{M}_{\alpha\beta}^{(3\times 3)}$.

In this talk, we will investigate this scenario in further detail.  We
will compute the mass eigenstates, masses and mixing angles and
investigate the question of CP violation, both in the quark and lepton
sector and possible connections between the two.

\section{Fermion Mass Matrices}

Our discussion is based on a six-dimensional (6D) GUT model
\cite{Asaka:2003iy}; for details, see
Refs.~\cite{Buchmuller:2004eg,Buchmuller:2007xv}.  The fermion mass
matrices have the following simple structure:
\begin{subequations}
  \label{eq:mass-matrices}
  \begin{align}
    \label{eq:mass-matrices-up}
    & \frac{1}{\tan{\beta}}\, m^u \sim \frac{v_u}{M_{B-L}}\ m^N \sim
    \begin{pmatrix}
      \mu_1 & 0 & 0 \cr 0 & \mu_2 & 0 \cr 0 & 0 & \mu_3
    \end{pmatrix}
    ,
    \\[2pt]
    \label{eq:mass-matrices-down}
    & m^d \sim m^e \sim \frac{1}{\tan{\beta}}\ m^\nu \sim
    \begin{pmatrix}
      \mu_1 & 0 & 0 & \tm_1 \cr 0 & \mu_2 & 0 & \tm_2 \cr 0 & 0 & \mu_3
      & \tm_3 \cr \tM_1 & \tM_2 & \tM_3 & \tM_4
    \end{pmatrix}
    ,
  \end{align}
\end{subequations}
where $\tan\beta=v_u/v_d$ is the ratio of the vacuum expectation values
(vevs) of the two Higgs doublets.  The hypothesis of a universal
strength of Yukawa couplings at each fixpoint leads to the
identification of the diagonal and off-diagonal elements of
$m^u/\tan{\beta}$, $m^d$, $m^e$ and $m^D/\tan{\beta}$ up to coefficients
of order one.  This implies an approximate top-bottom unification with
large $\tan{\beta}$ and a parameterization of quark and lepton mass
hierarchies in terms of the six parameters $\mu_1,\mu_2,\mu_3$ and
$\tm_1,\tm_2,\tm_3$.

The first three rows of the matrices are proportional to the weak scale.
The corresponding Yukawa couplings have to be hierarchical in order to
obtain a realistic spectrum of quark and lepton masses.  The fourth row
is of order the unification scale and, we assume, non-hierarchical.

The matrices in Eq.~(\ref{eq:mass-matrices-down}) can be diagonalized
using unitary matrices \mbox{$m = U_4 U_3 D\, V_3^\dagger V_4^\dagger$},
where $U_4$ and $V_4$ single out the heavy mass eigenstate, while $U_3$
and $V_3$ act only on the SM flavor indices and perform the final
diagonalization also in the $3\times 3$ subspace.

The parameters in the matrices (\ref{eq:mass-matrices}) are generally
complex.  With appropriate field redefinitions, however, $m^u$ is real
and we can absorb seven phases in $m$ and choose the remaining three
physical phases to be contained into the diagonal parameters $\mu_i$,
\begin{align} 
  \label{eq:m4-complex}
  m =
  \begin{pmatrix} 
    \abs{\mu_1} e^{i\theta_1} & 0 & 0 & \tm_1 \cr 
    0 & \abs{\mu_2} e^{i\theta_2} & 0 & \tm_2 \cr 
    0 & 0 & \abs{\mu_3} e^{i\theta_3} & \tm_3 \cr
    \rule[-0.5mm]{0mm}{5.5mm}\tM_1 & \tM_2 & \tM_3 & \tM_4
  \end{pmatrix} 
  .
\end{align}
With this choice, $V_4$ is real, while $U_4$ contains complex
parameters; however, the imaginary part is suppressed by
$\abs{\mu_i}/\tM $ so that their effect on the low energy CP violation
is negligible.  Then the discussion of the low energy CP violation
reduces to the case of three light generations.

The effective mass matrix is given by $\widehat{m}$, the $3\times 3$
part of
\begin{align}
  \label{eq:mhat}
  m' & = U_4^{\dagger} m V_4 =
  \begin{pmatrix}
    \widehat{m} & 0 \cr 0 & \tM
  \end{pmatrix} 
  + \mathcal{O}\left(\frac{v^2}{\tM^2}\right) .
\end{align}
As any matrix, $\widehat{m}$ can be transformed into upper triangular
form just by basis redefinition on the right,
\begin{align}
  \label{eq:triagonal-matrix}
  \ovl{m} & = \widehat{m}\ \widehat{V}_3 =
  \begin{pmatrix}
    \gamma\bm_1 & \bm_1 & \beta\bm_1 \cr 0 & \bm_2 &
    \alpha\bm_2 \cr 0 & 0 & \bm_3
  \end{pmatrix}
  \; .
\end{align}
The expressions for the parameters are given in
Ref.~\cite{Buchmuller:2007xv}.
We can choose $\bm_2$, $\bm_3$ and $\gamma\bm_1$ to be real, while
$\alpha$, $\beta$, and $\bm_1$ remain complex.

The form (\ref{eq:triagonal-matrix}) is particularly suitable for the
down quarks, where $\widehat{V}_3$ acts on the RH quarks and disappears
from the low energy Lagrangian due to the absence of RH current
interactions.  The matrix $\widehat{V}_3$ differs from the upper
$3\times 3$ part of the diagonalizing matrix $V_3 = \widehat{V}_3
V_3^\prime$; however, they are very similar in the hierarchical case.
The $3\times 3$ part of $U_3$ is the CKM matrix.

For the leptons, $V_4 V_3$ acts on the LH states, so the mismatch
between the charged lepton and neutrino basis appears in the charged
current interaction and the definition of the flavor neutrino
eigenstates.  However, the heavy state is an $\text{SU(2)}_L$ doublet,
so $V_4$ is identical for charged leptons and neutrinos.  The PMNS
matrix will only be given by the mismatch between the $V_3$ matrices for
charged leptons and neutrinos.

\section{Quark masses, mixing, and CP violation}

We can choose the parameters in such a way to give a consistent quark
mass pattern and CKM matrix \cite{Asaka:2003iy},
\begin{align} 
  & m_u : m_c : m_t \sim \mu_1 : \mu_2 : \mu_3 \; , \nonumber
  \\[2pt]
  & m_b = \bm_3 \; , \qquad m_s \sim \bm_2 \; , \qquad m_d \sim V_{us}
  \abs{\mu_2} , \nonumber
  \\
  & V_{us} \simeq \frac{\tm_1}{\tm_2} \;, \qquad V_{cb} \sim
  \frac{\tm_2}{ \tm_3} \; , \qquad V_{ub} \sim \frac{\tm_1 }{\tm_3}\; .
\end{align}
To describe CP violation for three generations, it is convenient to use
the Jarlskog invariant.  Since $H_u$ is diagonal in our model, the
invariant simply reads
\begin{align}  
  \label{eq:jarlskog}
  J_q & = \frac{\im \left( H_d^{12} H_d^{23} H_d^{31} \right)}{\Delta
    \mathscr{M}_d^2} \ .
\end{align}
where, due to the heaviness of the fourth generation, \mbox{$H_d =
  \widehat{m}\, \widehat{m}^{\dagger} = \ovl{m}\, \ovl{m}^{\dagger}$}
\cite{Buchmuller:2007xv} and
\begin{align}
  \Delta \mathscr{M}^2 = \left( m_3^2 - m_2^2 \right) \left( m_3^2 -
    m_1^2 \right) \left( m_2^2 - m_1^2 \right) .
\end{align}
Using Eq.~(\ref{eq:triagonal-matrix}), we obtain
\begin{align}
  J_q & = \frac{\bm_2 \bm_3^2}{\Delta \mathscr{M}_d^2} \im \left[
    \left(\alpha\bm_2\right) \left(\beta\bm_1\right)^\ast \bm_1 \right]
  .
\end{align}
The Jarlskog invariant is always independent of the argument of
$\gamma$.
It vanishes in the limit of vanishing down-quark mass, i.e., for
$\mu_1,\,\tm_1\to 0$ such that $\bm_1=0$ and for $\mu_1,\,\mu_2\to 0$,
which yields $\alpha=\beta$ (cf.~Eqs.~(\ref{eq:m4-complex}),
(\ref{eq:triagonal-matrix})).
Thus the presence of a single phase in $\alpha$ is not sufficient to
give CP violation in the low energy: this phase cancels out in the
Jarlskog invariant.  This effect stems from the alignment of the vectors
in flavor space; however, even in the case of vanishing first generation
mass, the corresponding eigenvector does not decouple from the other two
and the mixing matrix does not reduce to the two-generational case
\cite{Buchmuller:2007xv}.

Now, the down quark is not massless and the real physical case
corresponds to non-zero $\mu_1$, $\mu_2$ and $\tm_1$.  From the up quark
phenomenology, we know that $\mu_1:\mu_2$ is similar to the mass ratio
of up and charm-quark \cite{Asaka:2003iy}; in addition, $\tm_1 : \tm_2$
is fixed by the Cabibbo angle.  We will therefore focus on the linear
terms in $\mu_2$ and keep $\mu_1\simeq 0$.

Contributions to $J_q$ come from the complex quantities $\alpha\bm_2$,
$\beta\bm_1$, and $\bm_1$; however, $\beta\bm_1$ is independent of
$\mu_2$.
The first order terms give
\begin{align} 
  J_q & \simeq - \frac{\tm_1^2\tm_2^2\tm_3^2}{\Delta \mathscr{M}_d^2}
  \left[ \im\frac{\mu_3\mu_2^\ast}{\tm_3\tm_2} +
    \frac{\abs{\mu_3}^2}{\tm_3^2}\, \im\frac{\mu_2}{\tm_2} \right] .
\end{align}
We see that $J_q$ vanishes if either $\mu_2$ or $\mu_3$ vanish, so two
complex quantities are needed to obtain CP violation at low energies.

Substituting the order of magnitude of the parameters, we obtain
\begin{align}
  \label{eq:quark-phase-next}
  J_q & \simeq V_{us} \frac{m_d m_s}{m_b^2}\, \frac{1}{4\sqrt{2}} \left(
    3\, \sin\left(\theta_3-\theta_2\right) + \sin\theta_2 \right)
\end{align}
with $\tm_3 \simeq \abs{\mu_3}$.  The prefactor is of order $10^{-5}$,
which has to be compared with the current experimental value is
$J_q=3\times 10^{-5}$ \cite{Yao:2006px}.  Hence, at least one phase has
to be large.

\section{Lepton Mass Matrices}

The charged lepton and Dirac neutrino mass matrices can be transformed
like the down quark mass matrix.
Although charged lepton and down quark parameters are potentially
related by GUT symmetries, the corresponding phases after the
redefinition are completely uncorrelated.  Thus, there is no direct
relation between the CP violation in the leptonic and in the hadronic
observables.  We will see that different combinations of the phases
determine the experimental observables.  Hence, even if there were
relations between the phases in the quark and lepton sector, these would
not be observable.

The discussion of the charged lepton masses closely follows the
discussion of the down quarks.  The parameters are chosen such that they
match the observed hierarchy \cite{Buchmuller:2007xv}.  Here, we have to
take into account that the mass ratio of electron and muon is much
smaller than the ratio of down and strange quark.  This implies
$(\mu_2\tm_1/\tm_2^2)_e \ll (\mu_2\tm_1/\tm_2^2)_d$.  Assuming that the
difference is due to the smallest matrix elements, this indicates
$(\mu_2)_e/(\mu_2)_d \ll 1$ and/or $(\tm_1)_e/(\tm_1)_d \ll 1$ for
$(\tm_2)_e \simeq (\tm_2)_d$.  This fact can easily be accommodated in
the 6D model \cite{Asaka:2003iy}; for the general case, we have to allow
different values for the small parameters.

The light neutrino masses result from the seesaw mechanism, 
\begin{align}
  \label{eq:Majoranalight}
  m^\nu_\text{eff} & = - \left( m^D \right)^\top \left( m^N
  \right)^{-1} m^D .
\end{align}
The Majorana matrix is diagonal, but can have complex entries
\begin{align} 
  m^N = e^{2i \phi_3} \diag \left( M_1 e^{2i \Delta\phi_{13}}, M_2 e^{2i
      \Delta\phi_{23}}, M_3 \right) ,
\end{align}
where $\Delta\phi_{ij}=\phi_i-\phi_j$.  Altogether, we have nine
independent phases in the lepton sector; in the limit of small $\mu_1$
and $\rho_1$, they reduce to seven.  Since $\tM \sim M_\text{GUT}$ is
much larger than the Majorana masses $M_i$, we will neglect any effect
of this heavy fourth generation doublet and concentrate on the three
light generations including the RH neutrinos.

Since neither $\widehat{m}^e$ nor $\widehat{m}^D$ is diagonal, we will
change the basis in order to simplify the discussion of the CP
violation.  Thanks to the same hierarchical structure, the large
rotations of type $\widehat{V}_3$ (see Eq.~(\ref{eq:triagonal-matrix}))
are similar for charged leptons and neutrinos.  We use exactly the same
$\widehat{V}_3$ that transforms the charged lepton mass matrix into the
upper triangular form; the complete diagonalization can be obtained by
applying another nearly diagonal rotation matrix on the right,
corresponding to the mismatch between $V_3$ and $\widehat{V}_3$, and a
CKM-like rotation $U_3$ on the left.  Note that the transformations from
the left, $U_4$ and $U_3$, act on the RH fields and leave both
$H=\ovl{m}^\dagger\, \ovl{m}$ and the light neutrino Majorana mass
matrix, $m^\nu_\text{eff}$, unchanged.

Now we can write the effective Dirac neutrino mass matrix in the basis,
where the charged leptons are diagonal, as
\begin{align}  
  \label{eq:neutrino-dirac-triagonal}
  \ovl{m}^D & =
  \begin{pmatrix}
    A \rb_1 & D \rb_1 & \rb_1 \cr B \rb_2 & E \rb_2 & \rb_2 \cr C
    \rb_3 & F \rb_3 & \rb_3
  \end{pmatrix}
  .
\end{align}
Here we use $\rho_i$ and $\rt_i$ instead of $\mu_i$ and $\tm_i$ in order
to distinguish the parameters of $m^\nu$ from those in $m^e$.  Again,
the expressions for the parameters are given in
Ref.~\cite{Buchmuller:2007xv}.

We assume $\rb_3 : \rb_2 : \rb_1 \sim \rt_3 : \rt_2 : \rt_1 \sim m_b :
m_s : m_d$.  Then the light neutrino mass spectrum is hierarchical with
$m_1 \sim m_2 \sim \sqrt{\Delta m_\text{sol}^2}$, $m_3 \sim \sqrt{\Delta
  m_\text{atm}^2}$ and the weak hierarchy in the neutrino sector can be
traced back to the nearly perfect compensation between down and up quark
hierarchies.

Using the relations between $\rt_i$, $\rb_i$ and $\rho_i$, and $\tm_i$,
$\bm_i$ and $\mu_i$ due to the hierarchical structure of the mass
matrices in our model, one obtains the simple expressions,
\begin{align}
  \label{eq:rough}
  A & \sim C \sim \frac{\mu_2}{\bm_2} \ , & B & \sim
  \frac{\rho_2}{\rb_2} - \frac{\mu_2}{\bm_2} \ , & D & \sim E \sim F
  \sim 1 \ .
\end{align}
Due to the small electron mass, we expect $A$ and $C$ to be small.  Then
the mixing angles are given by
\begin{align}
  \label{eq:mixing-angles}
  & \tan \theta_{23} \simeq \abs{F} , \quad
  \tan \theta_{12} \sim \frac{\abs{B}}{\abs{E-F}} \sqrt{1+\abs{F}^2} \,,
  \\[3pt]
  & \sin\theta_{13} \sim \frac{C}{\sqrt{1+\abs{F}^2}} +
  \frac{B\left(EF+1\right)}{\left(1+\abs{F}^2\right)^{3/2}}
  \sqrt{\frac{\Delta m_\text{sol}^2}{\Delta m_\text{atm}^2}} \;.
  \nonumber
\end{align}
The atmospheric mixing angle $\theta_{23}$ is naturally large.
For $C \sim 0.1$, the reactor angle $\theta_{13}$ is close to the
current upper bound and the large solar mixing $\theta_{12}$ can be
achieved for $B \sim 0.1-1$ with moderate tuning of $E-F$.
For smaller values, $C \sim 0.01$, $\theta_{13}$ is dominated by the
second term in Eq.~(\ref{eq:mixing-angles}).  Then the angles
$\theta_{12}$ and $\theta_{13}$ depend on the same parameter $B$, which
should be relatively large.  Such value is not unnatural if we accept
$\rho_2 > (\mu_2)_e$.  In this case we have $\sin \theta_{13} \lesssim
0.1$ correlated with the mass eigenvalues $m_1 \lesssim m_2 \lesssim m_3
$.

The largest of the heavy neutrino masses is given by $M_3 \sim
m_t^2/\sqrt{\Delta m_\text{atm}^2} \sim 10^{15}\ {\rm GeV}$.  For the
lightest heavy Majorana state the model provides the rough estimate
$M_1 \sim M_3 m_u/m_t \sim 10^{10}\ {\rm GeV}$.

\section{CP violation in the leptonic sector}

\paragraph{Neutrinoless Double Beta Decay
  (\boldmath{$0\nu\beta\beta$}).}
The simultaneous decay of two neutrons may result in neutrinoless double
beta decay.  This process is currently most promising to prove the
Majorana nature of neutrinos.
The decay width can be expressed as
\begin{align}
  \Gamma & = G \abs{\mathcal{M}^2} \abs{m_{ee}}^2 ,
\end{align}
where $G$ is a phase space factor, $\mathcal{M}$ the nuclear
$0\nu\beta\beta$ matrix element, and $m_{ee}$ is the (11)-element of
the light neutrino mass matrix.

Since the electron mass is very small, $m^e$ in
Eq.~(\ref{eq:triagonal-matrix}) has nearly a vanishing first row.  Then
the LH electron is already singled out; the remaining rotation mostly
affects the (23)-block.  Therefore we can make an estimate of $m_{ee}$
from $m^\nu_\text{eff}$, which gives
\begin{align}
  \abs{m_{ee}} \sim \abs{\frac{\mu_2^2}{\bm_2^2} \sqrt{\Delta
      m_\text{atm}^2}\, e^{2i\Delta\phi_{23}} + \frac{\rho_2^2}{\rb_2^2}
    \sqrt{\Delta m_\text{sol}^2}} .
\end{align}
If $\mu_2/\tm_2 \ll \rho_2/\rt_2 $, the last term dominates, yielding
the familiar result for hierarchical neutrinos $\abs{m_{ee}} \lesssim
\sqrt{\Delta m_\text{sol}^2} \sim 0.01\ {\rm eV}$.

\paragraph{CP Violation in Neutrino Oscillations.}
Leptonic CP violation at low energies can be detected via neutrino
oscillations, which are sensitive to the Dirac phase of the light
neutrino mass matrix.  We can define a leptonic equivalent of the
Jarlskog invariant, which in our case simply reads
\begin{align}
  J_\ell & = -\frac{1}{\mathscr{M}_\nu^2}\, \im \left[
    \left(h^\nu_\text{eff} \right)^{12} \left(h^\nu_\text{eff}
    \right)^{23} \left(h^\nu_\text{eff} \right)^{31} \right] ,
\end{align}
with $h^\nu_\text{eff} = \left(m^\nu_\text{eff} \right)^\dagger
m^\nu_\text{eff}$.  
In general, the Dirac CP phase is given by a combination of the phases
of the neutrino Dirac mass coefficients $B$, $C$, $E$ and $F$ and no
useful upper bound on $J_\ell$ can be derived.

In the limit $\mu_2\to 0$, where $A=C=0$, but $B$ relatively large, the
leading term reads
\begin{align}
  J_\ell & \sim \abs{B}^2 \left( \kappa_1 - \kappa_2 \right) \ 
  \im\left(\Omega\right) ,
\end{align}
where \mbox{$\Omega = \left(1+EF^\ast\right) F^\ast \left(E-F\right)
  \frac{\varrho_2}{\varrho_3}$} and $\kappa_i$ are real parameters of
order unity \cite{Buchmuller:2007xv}.
The standard Dirac phase $\delta$ is suppressed by the ratio
$\abs{\varrho_2}/\abs{\varrho_3}$, as is $\sin\theta_{13}$.

\paragraph{Leptogenesis.}
The out-of-equilibrium decays of heavy Majorana neutrinos is a natural
source of the cosmological matter-antimatter asymmetry.  In recent years
this leptogenesis mechanism has been studied in great detail.  The main
ingredients are CP asymmetry and washout processes, which depend on
neutrino masses and mixings.

For hierarchical heavy neutrinos the baryon asymmetry is dominated by
decays of the lightest state $N_1$.  In our model, the corresponding CP
asymmetry reads
\begin{align}
  \varepsilon_1 & \simeq \frac{3}{8\pi} \frac{M_1}{v_u^2} \left(
    1-\frac{\tM_4^2}{\tM^2} \right)^{-1} \sum_{j=2,3}
  \frac{\rt_j^2}{M_j} \eta_j \ ,
\end{align}
where 
\begin{align}
  \eta_j & = - \im \left[ e^{i\Delta\phi_{j1}} \left( 1-
      \frac{\tM_4}{\tM} \frac{\rt_j \tM_4 + \rho_j^\ast \tM_j}{\rt_j
        \tM} \right)^2 \right] . \nonumber
\end{align}
The phases involved, $\Delta\phi_{13}, \Delta\phi_{12} $ and the phases
of $\rho_3,\rho_2$, are completely independent of the low-energy CP
violating phase in the quark sector and are not directly connected to
that in neutrino oscillations either (even if they can contribute to
it).
For $M_1\sim 10^{10}$\,GeV, one obtains $\varepsilon_1 \sim 10^{-6}$,
with a baryogenesis temperature $T_B \sim M_1 \sim 10^{10}\ \text{GeV}$.
These are typical parameters of thermal leptogenesis \cite{thermal}.

One finally obtains for the baryon asymmetry $\eta_B \sim 10^{-10}$,
consistent with observation.  So for successful leptogenesis we need a
non vanishing $\rt_1,\varrho_1 $ and in particular $\varrho_1 \sim
\varrho_2 $.  In such case a zero neutrino eigenvalue is only possible
due to alignment.

\section{Conclusions}

We have studied a specific pattern of quark and lepton mass matrices,
where up quarks and RH neutrinos have diagonal $3\times 3$ matrices with
the same hierarchy whereas down quarks, charged leptons and Dirac
neutrino mass terms are described by $4\times 4$ matrices which have one
large eigenvalue ${\cal O}(M_\text{GUT})$.  With this ansatz, the CKM
mixings are small because the LH down quarks barely mix with the extra
matter.  Conversely, RH down quarks and LH leptons have large mixings.

Neutrinos have a small mass hierarchy because of the seesaw mechanism
and the mass relations $m^d \sim m^\nu$ and $m^u \sim m^N$.  The
`squared' down-quark hierarchy is almost canceled by the larger up-quark
hierarchy.  The PMNS mixings are not suppressed by small mass ratios;
however, the neutrino phenomenology is largely fixed in terms of quark
masses and mixings.

The CP phases in the quark sector, neutrino oscillations and
leptogenesis are unrelated.  The measured CP violation in the quark
sector can be obtained, even if the CP invariant is suppressed by the
alignment between the two lightest mass eigenstates.  Quantitative
predictions for the lightest neutrino mass $m_1$ and $\theta_{13}$
require currently unknown $\mathcal{O}(1)$ factors.

The mass matrices (\ref{eq:mass-matrices}) can also arise in other
models, where additional matter is present at the GUT (or
compactification) scale.  Thus our results are generally valid for
models which share this matrix structure, independent of their origin.

\vspace*{-12pt}

\paragraph{Acknowledgments.}
It is a pleasure to thank W.~Buch\-m\"uller, L.~Covi and D.~Emmanuel-Costa
for collaboration and the organizers of SUSY07 for an enjoyable
conference.

\end{document}